\documentclass[aps,twocolumn,showpacs,amsmath,amssymb,superscriptaddress]{revtex4-1}

\usepackage{hyperref}
\usepackage{graphicx}
\usepackage{color}
\usepackage{dcolumn}
\usepackage{bm}
\usepackage{epsf}
\usepackage{amsmath,amstext,amssymb}
\usepackage{enumitem}
\usepackage[ansinew]{inputenc}
\usepackage{epstopdf}

\newcommand{\be}{\begin{equation}}
\newcommand{\ee}{\end{equation}}
\newcommand{\bea}{\begin{eqnarray}}
\newcommand{\eea}{\end{eqnarray}}
\newcommand{\bra}[1]{\langle #1\rvert}
\newcommand{\ket}[1]{\lvert#1\rangle}
\newcommand{\braket}[2]{\left\langle #1|#2\right\rangle}

\begin{document}

\title{Experimental realization of fast ion separation in segmented Paul traps}

\author{T.~Ruster}
\author{C.~Warschburger}
\author{H.~Kaufmann}
\author{C.~T.~Schmiegelow}
\author{A.~Walther}\altaffiliation{Present address: Department of Physics, Lund University, SE-22100 Lund, Sweden}
\author{M.~Hettrich}
\author{A.~Pfister}
\author{V.~Kaushal}
\author{F.~Schmidt-Kaler}
\author{U.~G.~Poschinger}

\affiliation{Institut f\"ur Physik, Universit\"at Mainz, Staudingerweg 7, 55128 Mainz, Germany}

\begin{abstract}
We experimentally demonstrate fast separation of a two-ion crystal in a microstructured segmented Paul trap. By the use of spectroscopic calibration routines for the electrostatic trap potentials, we achieve the required precise control of the ion trajectories near the \textit{critical point}, where the harmonic confinement by the external potential vanishes. The separation procedure can be controlled by three parameters: A static potential tilt, a voltage offset at the critical point, and the total duration of the process.  We show how to optimize the control parameters by measurements of ion distances, trap frequencies and the final motional excitation. At a separation duration of $80 \mu$s, we achieve a minimum mean excitation of $\bar{n} = 4.16(0.16)$ vibrational quanta per ion, which is consistent with the adiabatic limit given by our particular trap. We show that for fast separation times, oscillatory motion is excited, while a predominantly thermal state is obtained for long times. The presented technique does not rely on specific trap geometry parameters and can therefore be adopted for different segmented traps.
\end{abstract}


\maketitle

\section{Introduction}
The last decade has seen substantial progress in controlling the classical and quantum dynamics of trapped neutral atoms, ions and degenerate atomic gases in tunable double-well potentials. These systems offer the possibility to study and control the single- to double-well transition - being the paradigmatic instance of a critical phenomenon - in the quantum regime. For neutral atoms, confining potentials can be optical dipole potentials, for which control techniques are well established. This has for instance been used to study tunneling effects in tailored potential geometries \cite{folling2007direct,kierig2008single}. With degenerate quantum gases, fascinating experiments such as matter wave interferometry \cite{shin2004atom,schumm2005matter} or demonstration of a bosonic Josephson junction \cite{albiez2005direct} have been performed. For trapped ions, the confining potential can be dynamically controlled in micro-structured segmented Paul traps by application of suitable voltage waveforms to the trap segments. Proper control of single- to double-well transition enables the realization of separation and merging operations with Coulomb crystals. These operations are particularly interesting  in the context of the seminal proposal by Kielpinski, Monroe and Wineland \cite{KIELPINSKI2002} for the realization of scalable quantum information architectures with segmented ion traps. Besides laser- or microwave-driven logic gates, it is necessary to move ions within the trap in order to scale experiments up to a higher number of ions. These operations comprise shuttling of single or multiple ions and separation and merging of ion crystals \cite{ROWE2002}. Furthermore, crystal rotations \cite{splatt2009deterministic} can offer an alternative to swap gates, and more complex trap structures offer the possibility to shuttle ions around junctions \cite{hensinger2006tjunction,blakestad2009highfidelity}. To allow for the execution of subsequent logic gates, shuttling and separation operations have to be performed not only on a timescale much faster than the decoherence time of the system, but also have to feature a low energy increase. Other applications include remote coupling \cite{HARLANDER2010,brown2011coupled}, controlled interactions for quantum logic clocks \cite{rosenband2008frequency} or quantum simulation of bosonic many-particle systems \cite{lau2012proposal}.

\begin{figure}[t!]
	\centering
		\includegraphics[width=0.99\columnwidth]{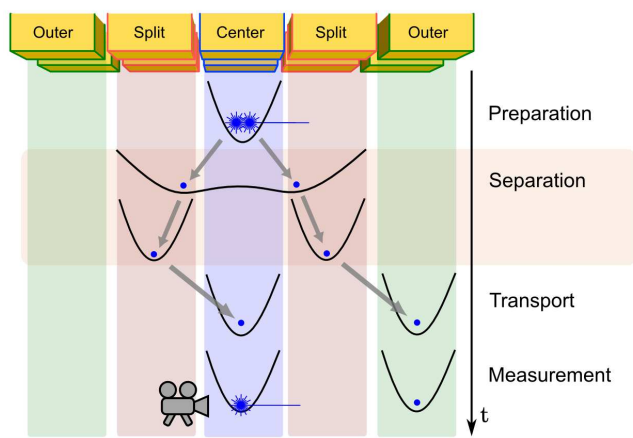}
	\caption{Schematic view of the relevant trap region and the measurement process. Time-dependent voltages are applied to the Outer($O$), Split($S$) and Center($C$) segments for controlling the shape of the axial electrostatic potential (black curves).  The ions are initialized at $C$. After separating the ion crystal, each ion resides at its respective $S$ segment. Then, after ion 2 is moved to its $O$ electrode, and ion 1 is transported back to $C$, where the motional state is measured.}
	\label{fig:sketch}
\end{figure}

For trapped ions in segmented traps, crystal separation is performed by supplying voltage waveforms to the dc trap segments. The common potential well at the initial electrode is lifted, while simultaneously two separate wells are created at the two neighboring electrodes, see Fig. \ref{fig:sketch}. First experimental realizations showed large amounts of excess energy transfer \cite{ROWE2002}. Since then, experimental techniques have been improved and separation has been used for demonstrations of e.g. deterministic quantum teleportation \cite{BARRETT2004}, entanglement purification \cite{reichle2006experimental} and a programmable two-qubit processor \cite{HANNEKE2010}. Recently, fast separation of a two-ion crystal that maintains a low energy increase at a duration as low as 55~$\mu$s has been reported \cite{BOWLER2012}. Compared to single-ion shuttling operations, which have been demonstrated at a few-$\mu$s timescale at negligible energy increase \cite{BOWLER2012,WALTHER2012}, separation represents a challenge for scalable experimental protocols. These operations show a strong sensitivity to imperfect control settings. The underlying reason is that the harmonic confinement transiently vanishes during the process, which makes the ions susceptible to energy increase mechanisms. Future scalable ion trap architectures require all operations to be much faster than the decoherence timescales. Thus, for the separation operations not to represent a major limitation, reliable optimization procedures are needed. In this paper we experimentally realize a separation protocol introduced in \cite{kaufmann2014robust}, which is largely independent of the trap geometry, and we demonstrate a universal procedure for optimizing the important parameters controlling the process. The manuscript is organized as follows: In Sec.\ref{sec:setup}, we describe relevant aspects of our ion trap setup along with basic measurement techniques, while we explain the used methods for energy measurements, trap calibration and voltage waveform design in Sec. \ref{sec:methods}. The results are presented in Sec. \ref{sec:results}.

\begin{figure*}[h!t]
	\centering
	\includegraphics[width=0.98\textwidth]{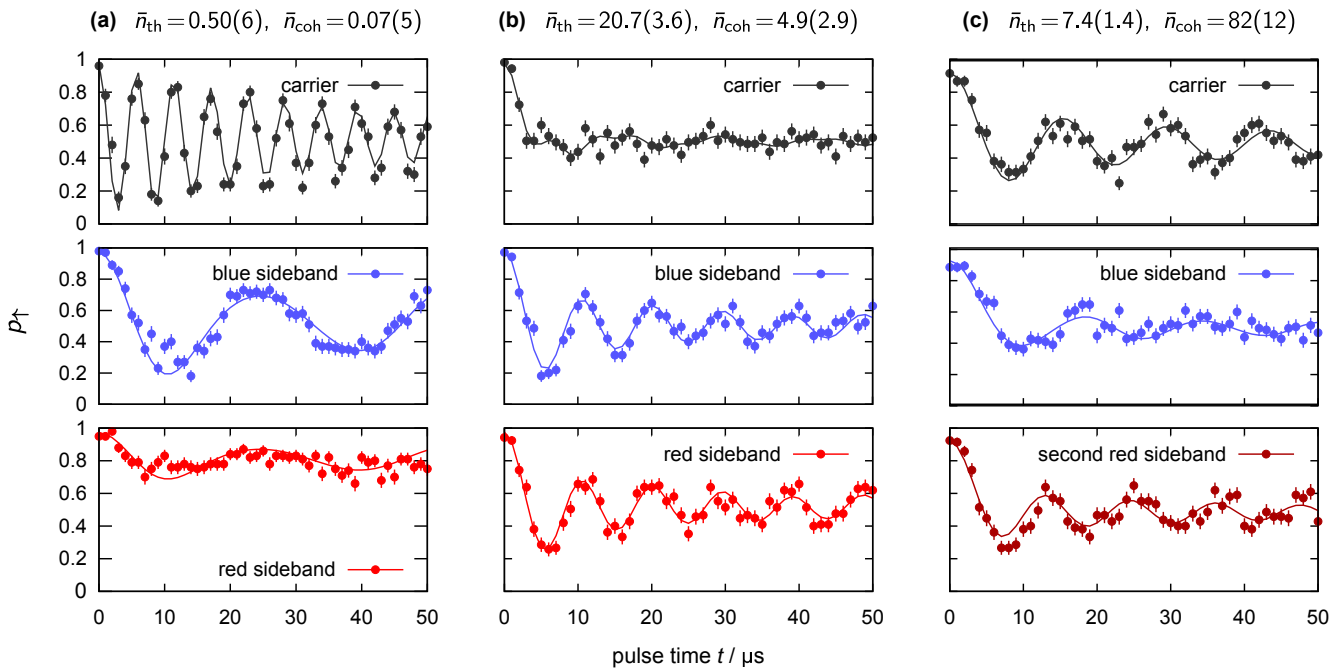}
	\caption{Examples for the analysis of the vibrational state of a single ion by Raman excitation: Column \textbf{(a)} shows Rabi oscillations for an ion cooled close to the motional ground state, with (top to bottom) the carrier transition ($\Delta n=0$), blue sideband ($\Delta n=+1$) and red sideband ($\Delta n=-1$). Blue and red sideband clearly exhibit strongly different signals, a characteristic of the Lamb-Dicke regime. \textbf{(b)} shows similar data for an ion with thermal excitation of $\bar{n}_{th}=$20.7(3.6), where no coherent dynamics is observed on the carrier, while the sidebands display Rabi oscillations. \textbf{(c)} shows data for a strong oscillatory excitation $\vert\alpha\vert^2=$~82(12), where the second red sideband (bottom) is used rather than the first in order to gain more information from the measurement. In all panels, the solid lines originate from a simultaneous fit of the data pertaining to the different transitions.}
	\label{fig:figureRabi}
\end{figure*}

\section{Experimental setup}
\label{sec:setup}

We store $^{40}$Ca$^+$ ions in a micro-structured segmented Paul trap similar to \cite{SCHULZ2006,SCHULZ2008}. The electrode width is 200$\mu$m \footnote{The complete trap dimensions are given in Ref. \cite{kaufmann2014robust}, Sec. 3, Trap A}, and an axial trap frequency of about $(2\pi)\cdot$1.4~MHz is obtained for a trapping dc voltage of -7~V at segment $C$, with the other segments at 0~V. The ions are tightly confined in the radial directions, with radial trap frequencies in the range of $2\pi\cdot$2.5-4~MHz. In the following, we consider only motion along the trap axis . A two-ion crystal initially stored at the center electrode $C$ (Figure \ref{fig:sketch}a), where the ions are also detected, is to be split in two ions trapped in separate potential wells at the neighboring split electrodes $S$. The outer electrodes $O$ are used to null a possible background force along the trap axis, and they provide additional axial confinement during the separation. 

The dc trap electrodes are supplied with time-dependent voltages, which are generated by a custom-made FPGA-based arbitrary waveform generator \cite{WALTHER2012}. It allows for output voltages in the range $\pm$10~V with a resolution of about $0.3$~mV and a maximum update rate of $2.5$~MSamples/s. Similar hardware has also been used in \cite{BAIG2013, BOWLER2013}. The waveforms have to be filtered with a sufficiently strong noise suppression at the minimum attained trap frequency. We employ second-order $\Pi$-type low-pass filters with a cutoff frequency of 50~kHz for each segment to suppress noise arising from voltage updates. 

Doppler cooling of the ions is accomplished by illumination on the  $S_{1/2} \leftrightarrow P_{1/2}$ cycling transition near 397~nm. Resonance fluorescence is detected on an EMCCD camera \footnote{Andor iXon 860 at a magnification of about 57}.
The distance between trapped ions is measured by imaging the ions' resonance fluorescence \cite{jechow2011wavelength}. From the acquired image, the distance in EMCCD pixels is extracted. The spatial distance is obtained by rescaling using calibration data of the imaging system. 
The magnification of the imaging system is determined for a harmonically trapped two ion crystal (mass $m$, charge $q$). The observed distance is given by $d=\left(2\pi \epsilon_0 m\omega^2\right/q^2)^{-1/3}$, where $\omega$ denotes the center-of-mass frequency, which is experimentally determined from resolved sideband spectroscopy. This method enables us to determine ion distances with an accuracy of about 0.1~$\mu$m, corresponding to about 1/4 of an EMCCD pixel.\\
For ground-state cooling and measuring the motional excitation, we use two perpendicularly propagating, off-resonant laser beams to drive stimulated Raman transitions between the ${\ket{m_J = +1/2}\equiv\ket{\uparrow}}$ and ${\ket{m_J = -1/2}\equiv\ket{\downarrow}}$ sublevels of the $S_{1/2}$ ground state \cite{POSCHINGER2009,WALTHER2012}. These spin states are split by approximately 10~MHz by an external magnetic field. The beam geometry gives rise to a coupling only to the axial mode of vibration, characterized by a Lamb-Dicke factor of $\eta \approx 0.23$.  
The spin is read out by spin-selective electron shelving on the $S_{1/2} \leftrightarrow D_{5/2}$ quadrupole transition near 729~nm, where only population from $\ket{\uparrow}$ is transferred to the metastable state $D_{5/2}$ by means of rapid adiabatic passage \cite{POSCHINGER2009}. Subsequent illumination on the cycling transition yields resonance fluorescence if the ion is projected to the $\ket{\downarrow}$ state. The quadrupole transition is also employed for precise measurements of secular trap frequencies via resolved sideband spectroscopy, where an accuracy of about 2~kHz is readily obtained by measuring the frequency difference between carrier transition and blue motional sideband.

\section{Methods}
\label{sec:methods}
\subsection{Measurement of motional excitation}
\label{sec:measscheme}

Each experimental run is started by Doppler cooling a two-ion crystal on the $C$ segment, where in all laser-ion interactions take place. We then apply resolved sideband cooling of both axial modes close to the motional ground state, reaching mean phonon numbers of about $\bar n_{\text{COM}} \approx 0.7$ and $\bar n_{\text{STR}} \approx 0.7$. After optically pumping both ions to the state $\ket{\uparrow}$, a sequence of separation and transport operations is executed (Figure \ref{fig:sketch}b). First, the crystal is separated such that both ions move to the respective $S$ segments. One ion is then adiabatically transported further away to the neighboring $O$ segment within a shuttling time of 24~$\mu$s and kept there. The other ion is adiabatically transported back to $C$, where its energy increase is measured by driving the stimulated Raman transition and subsequent measurement of the qubit state. To restore the initial situation after detecting the state of the ion, the sequence of transports and separation operations is applied in reverse order. The sequence can be carried out with mirrored transports to measure the state of the second ion.\\
For the energy increase measurement, we drive the stimulated Raman transition on the carrier or on the axial sidebands with a pulse time $t$, after the sequence of shuttling operations is completed. Subsequent spin readout reveals the occupation probability $p_{\uparrow,\Delta n}(t)$ of being in the state $\ket{\uparrow}$ after driving the transition corresponding to a phonon number change $\Delta n$\cite{LEIBFRIED2003}. The osccupation probabilities are inferred by repeating a measurement pertaining to fixed $\Delta n, t$ by 200 times. By scanning $t$ and acquiring data for several $\Delta n$, we can infer the phonon probability distribution $p_n$ by using the relation
\begin{equation}
	p_{\uparrow,\Delta n}(t) = \sum_{n=0}^{\infty} p_n \sin^2 \left( \Omega M_{n,\Delta n} t/2 \right)
\end{equation}
with the bare Rabi frequency $\Omega$ and the transition matrix elements $M_{n,\Delta n}$ \cite{LEIBFRIED2003} corresponding to the specific transition $\Delta n$. For mean phonon numbers $\overline n \leq 20$, we probe the carrier transition ($\Delta n=0$) as well as the red and blue sidebands ($\Delta n = \pm 1$). For higher phonon numbers, we also probe the second red sideband ($\Delta n =-2$). 
The data is well described by a phonon number distribution characterized by a coherent displacement $\zeta$ and a thermal mean phonon number $\bar{n}_{\text{th}}$:
\begin{equation}
p_n=\bra{n} \hat{D}^{\dagger}(\zeta)\hat{\rho}_{\text{th}}\hat{D}(\zeta)\ket{n},
\label{eq:pn}
\end{equation}
with the thermal distribution $\langle n\vert \hat{\rho}_{\text{th}}\vert n\rangle=\bar{n}_{\text{th}}^n/(\bar{n}_{\text{th}}+1)^{(n+1)}$ and the displacement operator $\hat D(\zeta)$, where the displacement amplitude $\zeta$ corresponds to the mean phonon number $\bar{n}_{\text{coh}}=\vert\zeta\vert^2$. Values for $p_n$ can be obtained directly by evaluating Eq. \ref{eq:pn} for small phonon numbers $n\leq 20$. 
\footnote{Note that indicate the motional excitation in phonons pertaining to a single $^{40}$Ca$^+$ ion in a harmonic potential at a frequency of $2\pi\cdot$1.4~MHz, even if we are dealing with large thermal or oscillatory excitations far from the quantum regime. The usage of phonons as energy unit makes it convenient to compare the outcome with the requirements for performing gate operations.}
For larger phonon numbers, the phonon distribution is obtained by numerical thermalization of the phonon distribution corresponding to a displaced state, see \ref{sec:thermaldisplaced}.\\
Data for several transitions $\Delta n$ is jointly used to infer the thermal and oscillatory mean phonon numbers $\bar{n}_{\text{th}}$ and $\bar{n}_{\text{coh}}$ by performing a Bayesian parameter estimation based on a Markov chain Monte-Carlo sampling method. This method enables us to infer excitations of up to $\bar{n}_{\text{coh}} \lesssim 400$ at relative accuracies of about $5\%$ near the ground state to $10\%$ for large excitations, and allows for distinguishing  thermal and oscillatory excitation. Case examples for Rabi oscillation data are shown in Fig. \ref{fig:figureRabi}.

\subsection{Trap characterization}
We parametrize the electrostatic potential along the trap axis $x$ by a Taylor approximation around the initial center-of-mass position at $C$ \cite{HOME2006}
\begin{equation}
	V(x,t)\approx\beta(t)~x^4+\alpha(t)~x^2+\gamma(t)~x.
	\label{eq:taylorpot}
\end{equation}
This approximation holds for ion distances much smaller than the width of a trap segment. 
To keep the notation uncluttered, we omit explicit time dependence in the following. The coefficients are determined by the time-dependent electrode voltages $U_{C,S,O}$ and the trap geometry:
\begin{align}
	\alpha &= U_C\alpha_C+U_S\alpha_S+U_O\alpha_O+\alpha' \label{eq:alphadef} \\
	\beta &= U_C\beta_C+U_S\beta_S+U_O\beta_O+\beta' \label{eq:betadef} \\
	\gamma &= \Delta U_S\gamma_S+\Delta U_O\gamma_O+\gamma'\label{eq:gammadef}
\end{align}
The voltage $U_S$ ($U_O$) is applied to both $S$ ($O$) electrodes, while the voltage $\Delta U_S$ ($\Delta U_O$) is a differential voltage between the electrodes comprising the respective pair. The constant coefficients $\alpha_i, \beta_i, \gamma_i$ are determined by the second, fourth and first derivatives of the respective normal electrode potentials at $x=0$. Experimental imperfections are characterized by offset coefficients $\alpha'$, $\beta'$, $\gamma'$. Such  contributions arise from stray charges, residual ponderomotive forces and asymmetries in the trap geometry. The values for the geometry coefficients could be in principle obtained by electrostatic simulations of the trap \cite{SINGER2010}. However, we find substantial deviations of the experimental values, such that the required degree of control for separation is not achieved.\\
The initial situation is characterized predominantly harmonic confinement $\alpha \gg 0$, and the trap frequency is given by $\omega^2=2e\alpha/m$. Separation of the two-ion crystal is  performed by sweeping $\alpha$ from a positive to a negative value, which transforms a single well potential at $C$ to a double well potential at the $S$ segments. At the \textit{critical point} (CP) $\alpha=0$, the axial confinement is at its minimum strength, while the rate of change of the equilibrium ion distance attains its maximum. This corresponds to a strong impulsive drag, possibly leading to oscillatory excitation of up 10$^6$ phonons \cite{kaufmann2014robust}. Thus, the voltage ramps need to be designed such that $\alpha$ is slowly varying close to the CP, which in turn requires the precise knowledge of coefficients $\alpha_i$ from Eq. \ref{eq:alphadef}. \\
For calibration of the $\alpha_i$, we apply resolved sideband spectroscopy to determine the axial trap frequency. Before the spectroscopy pulse, we ramp the segment voltages to a desired voltage configuration and then restore the initial voltages after the spectroscopy pulse has been applied. For determining $\alpha_C$ and $\alpha'$, we measure the axial trap frequency for a set of 6 different voltages $U_C$, while keeping the other voltages constant. Similarly, $\alpha_{S}$ ($\alpha_{O}$) is determined by setting 6 different values for $U_S$ ($U_O$). The $\alpha$ coefficients are then obtained by performing linear regression of the squared secular frequencies versus the corresponding voltage. This procedure enables us to measure all $\alpha$-parameters, including the imperfection $\alpha'$, with a sub-percent accuracy. For the specific set of trap segments we used for the measurements for this work, we find $\alpha_C=$~-2.612(7)$\cdot$10$^6$~m$^{-2}$, $\alpha_S=$~-1.279(12)$\cdot$10$^6$~m$^{-2}$, $\alpha_O=$~0.993(5)$\cdot$10$^6$~m$^{-2}$ and $\alpha'=$~-1.956(35)$\cdot$10$^6$~V m$^{-2}$. Note the substantial value of the offset parameter $\alpha'$, which would correspond to a spurious voltage of about +0.75~V applied to segment C. The values for the quartic potential parameters $\beta_i$ are calibrated by measuring ion distances in a purely quartic potential: We use the calibrated $\alpha_i$ values to apply voltage sets to the segments which correspond to the CP, $\alpha=0$. The ion distance is then given by $d_{eq}^{(CP)}=\left(2\kappa/\beta\right)^{1/5}$ \cite{kaufmann2014robust}, with $\kappa=e/4\pi\epsilon_0$ characterizing the Coulomb repulsion. Thus, measurement of the ion distance on the EMCCD camera reveals $\beta$, and performing this measurement for varying voltages $U_i$ reveals the segment coefficients $\beta_i$. For 22 different voltage sets, we find CP ion distances varying between 25~$\mu$m and 55~$\mu$m. We determine the quartic coefficients for the $C$ and $S$ segments to be $\beta_C=$~3.1(1)$\cdot$10$^{13}$~m$^{-4}$ and $\beta_S=$~-6.2(3)$\cdot$10$^{12}$~m$^{-4}$. Due to the small feed-through of the $O$ segments, no value for $\beta_O$ can be determined with this method. A large offset quartic coefficient $\beta'=$~1.5(1)$\cdot$10$^{14}$~ V m$^{-4}$ is found, indicating the presence of strongly inhomogeneous static background fields. Note that the contribution from $\beta'$  to the quartic potential at the CP is about 20\%.

\subsection{Tilt calibration}
\label{sec:tiltcalibration}
A residual electric field $\gamma'$ along the trap axis breaks the symmetry of the electrostatic potential. It can be sufficiently strong to tilt the potential at the CP such that the ions are not separated, but rather stay confined in one of the potential wells. The critical tilt value $\gamma_{\text{crit}}$ can be measured by monitoring slowly separating ions on a camera for different compensation voltages $\Delta U_O$. We find the separation to work within the window of compensation voltages $\Delta U_O^{(l)} \leq \Delta U_O \leq \Delta U_O^{(u)}$. The window center $\bigl(\Delta U_O^{(l)}+\Delta U_O^{(u)}\bigr)/2$ is assumed to roughly correspond to a canceled background field $\gamma=0$, while the half width $\delta U_O=\left(\Delta U_O^{(u)}-\Delta U_O^{(l)}\right)/2$ yields the critical tilt field via $\gamma_{\text{crit}}=\gamma_O\delta U_O$. Experimentally, we find  $\delta U_O\approx$~16 mV, corresponding to a force on a single ion at $x=0$ of only about 800~zepto Newton, underlining the strong sensitivity of the separation process on precise control of the trap potentials. The measurement accuracy for this tilt compensation scheme is limited by the resolution at which the segment voltages can be set, and the electric feedthrough of the segments used for the compensation. In our case, the voltage resolution is 0.3~mV, leading to a tilt field resolution of about 0.1~V/m, corresponding to a force resolution 15~zeptoN. In \cite{kaufmann2014robust}, it is shown that merely compensating to values within the window where the separation works is not sufficient, as nonzero tilt fields can lead to strongly increased motional excitation due to quasi-discontinuous trajectories of the equilibrium positions. Thus, we need to suppress, characterize and compensate for temporal drifts of the tilt field, which is explained in detail in Appendix \ref{sec:drifts}.

\subsection{Voltage ramp design}
The voltage ramps $\left\{U(t)\right\}= \left\{U_C(t), U_S(t), U_O(t)\right\}$ are designed in a two-step scheme. We exploit the fact that the harmonic coefficient $\alpha$ is monotonically decreasing throughout the separation process to determine voltage sets $U_i(\alpha)$ as a function of $\alpha$. First, we specify the voltages $U_S$ and $U_O$ at the start, at the CP and at the end of the separation process, and linearly interpolate between these values. The voltage $U_{C}$ is then obtained for each $\alpha$ by using Eq. \ref{eq:alphadef} and the measured trap calibration data:
\begin{equation}
U_C(\alpha)=\frac{1}{\alpha_C}\left(\alpha-\alpha'-\alpha_O U_O(\alpha)-\alpha_S U_S(\alpha)\right)
\label{eq:UCfromalpha}
\end{equation}
The initial voltage set $\{U^{(i)}\} = \{-7V, 0V, 0V\}$ gives rise to a trap with an axial frequency of about $\omega \approx 2\pi\cdot$~1.4~MHz. The voltages are ramped towards the CP corresponding to $\{U^{(CP)}\} \approx \{-1.89, -7.5V, +9V\}$, leading to a minimum axial frequency of about $\omega_{\text{crit}} \approx 2 \pi \cdot 170$~kHz. The final voltage configuration is specified to be $\{U^{(f)}\} = \{+2.62V, -7.83V, 0V\}$, leading to separate traps at the $S$ segments such that the axial trap frequency matches the initial one. A more refined scheme uses two additional mesh points near the initial(final) steps to ramp $U_O$ up(down), such that $U_O$ is constant around the CP.

\begin{figure*}[htp]
	\begin{minipage}[t]{0.49\textwidth} 
		\centering
                \includegraphics[width=\textwidth]{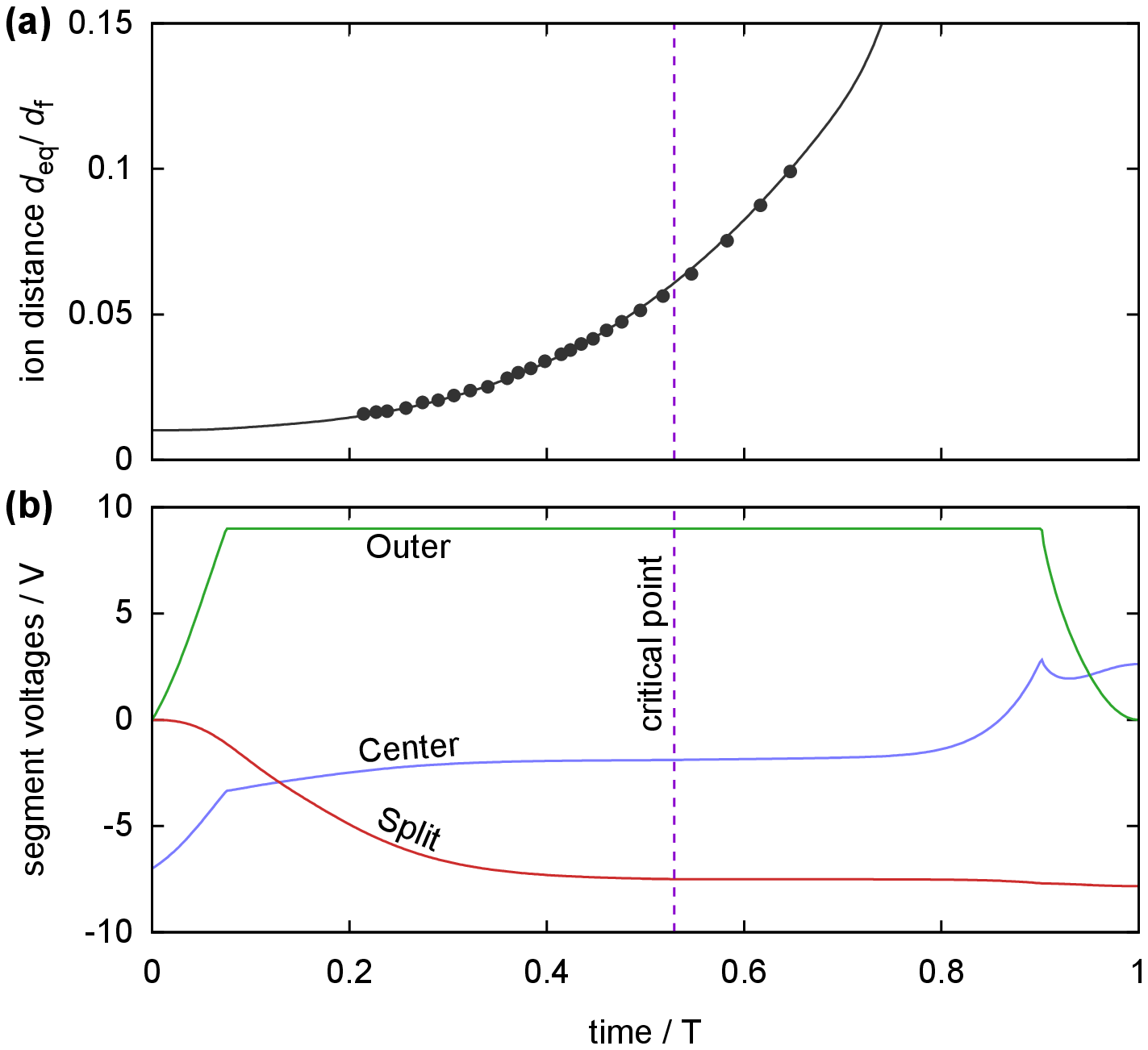}
                \caption{Time dependent distance and voltages. \textbf{(a)} Variation of the equilibrium distance $d_{eq}$ with time as given by Eq. \ref{eq:traj} and the truncation procedure (see text). The dots indicate distance measurements with an EMCCD camera and resolved sideband spectroscopy gauge. (b) Variation of the segment voltages during the separation process. In both panels, the CP is indicated by the vertical dashed line. Note the small variation of the voltages around the CP.}
                \label{fig:traj}
	\end{minipage}
	\hfill
	\begin{minipage}[t]{0.49\textwidth} 
		\centering
                \includegraphics[width=\textwidth]{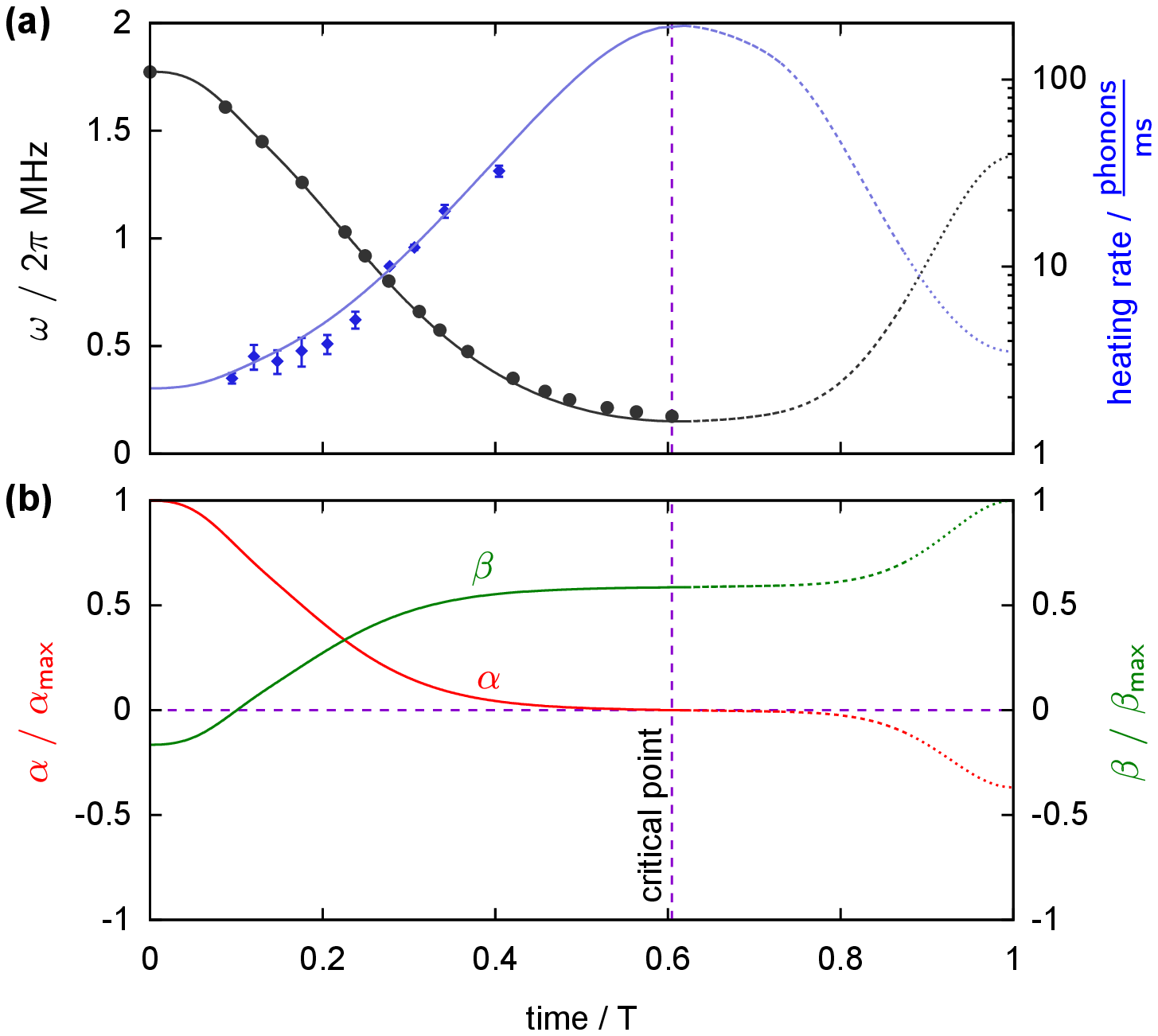}
                \caption{Time dependent trap frequency and potential coefficients. \textbf{(a)} shows the secular frequency pertaining to the COM vibration as calculated from Eq. \ref{eq:omegavsd} (black solid), along with resolved sideband spectroscopy measurements (black dots). From the trap frequency, we infer the anomalous heating rate (blue solid), where a power law is determined by heating rate measurements (blue dots). We obtain a minimum value of $\omega_{CP}\approx 2\pi\cdot$174(10)~kHz. \textbf{(b)} shows the variation of the harmonic ($\alpha$) and quartic coefficients ($\beta$), as obtained from trap calibration data and voltage ramps. Note the small variation of $\alpha$ around the CP. The quantities beyond the CP are shown dashed, as the Taylor approximation Eq. \ref{eq:taylorpot} breaks down and the precise equilibrium positions deviate.}
                \label{fig:omegaVsAlpha}
	\end{minipage}
\end{figure*}

In the second step, we calculate time-dependent voltage ramps $\{U(t)\}$ from the parametrized ramps $\{U(\alpha)\}$. This requires knowledge of the relation $\alpha \leftrightarrow d_{eq}$, where $d_{eq}$ is the equilibrium distance between both ions. We can calculate $d_{eq}$ for two ions in the potential  Eq. \ref{eq:taylorpot} by using the $\alpha$ and $\beta$ coefficients obtained from the calibration measurements and taking Coulomb repulsion into account. Based on this, is possible to design a trajectory $d_{eq}(t)$.\\ 
The choice for the distance function
\begin{equation}
	d_{eq}(t) = d_i+(d_f-d_i) \left(\frac{t}{T}\right)^2 \sin^2 \left(\frac{\pi}{2} \frac{t}{T}\right) 
\label{eq:traj}
\end{equation}
with initial (final) distance $d_i$($d_f)$ and final time $T$ fulfills the requirement of a small variation rate of $\alpha$ at the CP \cite{kaufmann2014robust}.  We further optimize by truncating the initial 10\% and final 30\% of $T$ by rescaling, leaving the CP voltages fixed. We verified experimentally that this does not affect the resulting motional excitations significantly, while the duration of the separation process becomes shorter.
We can finally calculate time dependent voltages $\{U(t)\}$ by i) determining the $d_{eq}(t)$ for a given time $t$ from Eq. \ref{eq:traj}, ii) finding the value of $\alpha$ for this distance, and iii) looking up the voltages for this $\alpha$ value from the linear interpolation ramps:
\begin{equation}
	\{U(t)\}: t \xrightarrow{i)}  d_{eq}(t) \xrightarrow{ii)} \alpha(d_{eq}(t)) \xrightarrow{iii)} \{U(\alpha(d_{eq}(t)))\}
\end{equation}	
The voltage ramps calculated with this scheme exhibit small variation rates around the CP. A example trajectory and the resulting voltage ramps are shown in Fig. \ref{fig:traj}. For fine-tuning the trajectory near the CP and for possible compensation of errors in the calibration measurements and distortions from the low-pass filters, we add an additional voltage offset $\delta U_C^{(CP)}$ to the $C$ segment, which is applied at the CP.

\section{Results}
\label{sec:results}

We first verify the precise control over the ion distance by measurements. A voltage set pertaining to a given equilibrium distance is applied to the trap segments, while the ions are continuously Doppler cooled. The ion distance is measured, see Sec. \ref{sec:setup}. A comparison of measured distances with the preset values determined from Eq. \ref{eq:traj} is shown in Fig. \ref{fig:traj}. For the accessible range of distances of up to roughly $2\cdot d_{CP}$, we find an agreement consistent with the accuracy of the distance measurement scheme.\\

We also characterize the weakening of the axial confinement when the CP is approached. This is done by dynamically separating two ions (without Doppler cooling) up to a fixed distance $d\leq d_{CP}$, corresponding to a fraction of the total separation time $T$. This is followed by the application of a spectroscopy pulse on the quadrupole transition, and moving the ions back to the initial distance before fluorescence readout. This way, we ensure that the determination of the secular frequency is not hampered by insufficient Doppler cooling at low trap frequencies \cite{poulsen2012efficient}. The secular frequency measured this way is a \textit{local} trap frequency given by the second derivative of the external potential at the position of each ion. It is given by \cite{HOME2006}:
\be
\omega^2(t)=\frac{q}{m}\left(3\beta(t) d(t)^2+2\alpha(t)\right).
\label{eq:omegavsd}
\ee
This corresponds to the COM secular frequency in the limit where the ions are confined in a common strongly harmonic potential well. The coefficients $\alpha,\beta$ are obtained by using Eqs. \ref{eq:alphadef},\ref{eq:betadef} with the voltage ramps in conjunction with the trap calibration data. Measurement results below the CP are shown in Fig. \ref{fig:omegaVsAlpha}. The data is matching well to Eq. \ref{eq:omegavsd}  \footnote{Note that the data shown in Fig. \ref{fig:omegaVsAlpha} is pertaining to slightly different voltage ramps than the other results shown in the manuscript, where the $O$ electrodes were constantly kept at the maximum voltage.}, and a CP trap frequency of about 10\% of the initial trap frequency is found. Such substantial reduction of the trap frequency leads to strongly increased anomalous heating rates\cite{NIZAMANI2012}. This scaling behavior is determined by adiabatically lowering the confinement for a single ground-state cooled ion by ramping up the trap voltage at segment $C$. The ion is kept at the lowered confinement for a given wait time before raising the confinement again to the initial value and probing the energy increase with the method described in Sec. \ref{sec:measscheme}. From linear regression of the energy increase with respect to the wait time, the heating rate at the trap frequency corresponding to a given lower confinement voltage is extracted. This way, we determine a phonon increase rate of $\Gamma(\omega)=6.3(2)\cdot\omega^{-1.8(1)}$~ms$^{-1}$. An estimate of the total thermal energy gain during the separation is obtained by integrating over the heating rate, which is time-dependent via the time-dependent trap frequency, i.e. the area under the heating rate curve of Fig. \ref{fig:omegaVsAlpha} a). As a result, we expect a thermal energy gain of 29(7)~ms$^{-1}\cdot T$ phonons per ion at a separation time $T$.\\
\begin{figure*}[htp]
	\centering
	\includegraphics[width=0.99\textwidth]{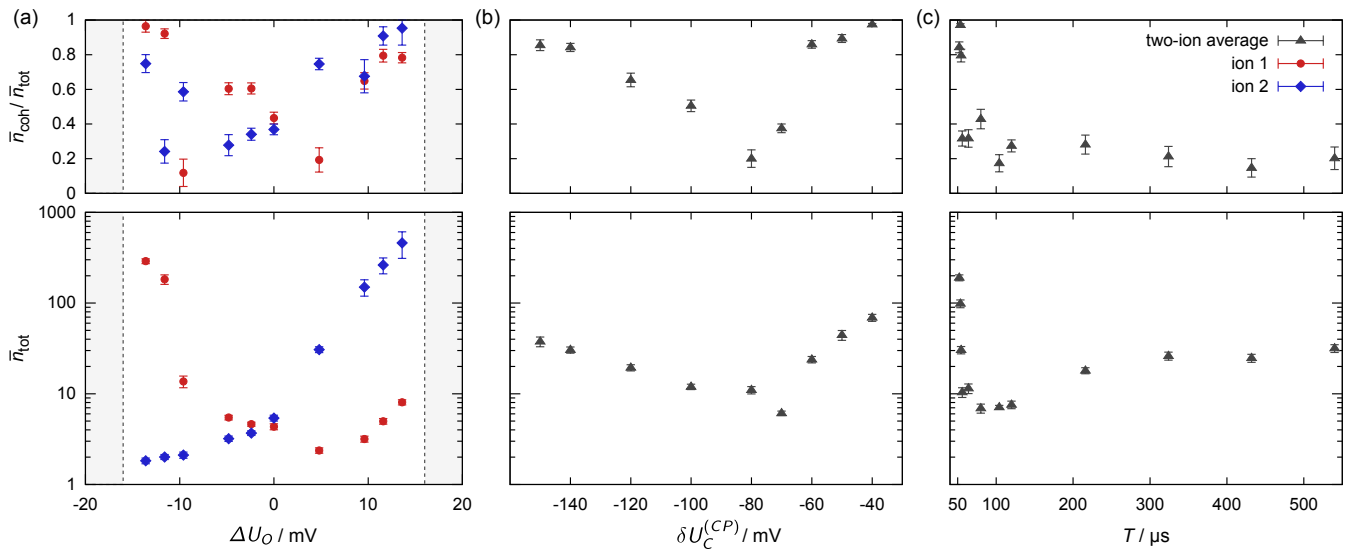}
	\caption{Measured total motional excitation and coherent fraction after the separation process versus (a) the tilt compensation voltage $\Delta U_{O}$ relative to the center of the separation success range, (b) the offset voltage of the $C$ segment at the CP $\delta U_C^{(CP)}$, and (c) the total separation duration $T$.\\
	For the dependence on $\Delta U_{O}$, we show the motional excitation for each ion, while for the other parameters the averaged excitation over both ions is shown.}
	\label{fig:resultsplot}
\end{figure*}

We investigate the dependence of the final motional excitation of the ions on three parameters: (i) the voltage $\Delta U_O$ modifying the tilt field $\gamma$, (ii) the CP offset voltage at the $C$ segment $\delta U_C^{(CP)}$, and (iii) the total separation duration $T$. The measurement results are shown in Figure \ref{fig:resultsplot}. For these measurements, a two-ion crystal, initially cooled to mean phonon numbers of $\bar{n}_{\text{COM}}\approx$0.7 on the center-of-mass mode and $\bar{n}_{\text{STR}}\approx$0.7 on the stretch mode, is separated. For probing the energy of one of the ions, a sequence of shuttling operations is carried out as indicated in Fig. \ref{fig:sketch}, where each shuttling to a neighboring segment is performed within 24~$\mu$s and using 60 samples, such that no significant energy transfer takes place \cite{WALTHER2012}. The energy is probed according to Sec. \ref{sec:measscheme}, such that the energy increase for each ion in phonons pertaining to the final single-ion trap frequency at $C$ is obtained.

The dependence on the tilt field $\gamma$ is characterized by measuring the motional excitation for different voltages $\Delta U_O$, for a total separation time of 80~$\mu$s. The results are shown in Fig. \ref{fig:resultsplot} a). We find that the energy increase of both ions is minimized  for a tilt field where the energy increase is roughly equal for the two ions. As the best result, we obtain $\bar{n}_{1,\text{tot}} = 4.63 \pm 0.23$ and $\bar{n}_{2,\text{tot}} = 3.69 \pm 0.22$. This voltage slightly differs from the center of the success range by about $-2$~mV. For deviation from the optimum voltage of about $4$~mV, strong oscillatory excitation on at least one of the ions corresponding to $>$10 phonons can occur. The dependence on the CP voltage offset $\delta U_C^{(CP)}$, as shown in Fig. \ref{fig:resultsplot} b), is less pronounced, it has to be correctly set to an accuracy of about $10$~mV to circumvent coherent excitation. However, a significant deviation of the optimum value of -70~mV with respect to the ideal case (0V) is observed, which presumably compensated for inaccuracies of the trap calibration data and voltage waveform distortion by the filters. For correct settings of $\delta U_C^{(CP)}$ and $\Delta U_O$, the motional excitation is dominated by heating.\\
Finally, we vary the total separation duration $T$, see Fig. \ref{fig:resultsplot} c). For each duration $T$, the offset voltage $\delta U_C^{(CP)}$ and the tilt compensation voltage $\Delta U_O$ are optimized separately in order to compensate for voltage waveform distortions which are caused by the 50~kHz low-pass filters. This is done efficiently by probing the carrier Rabi oscillation on the stimulated Raman transition at a $\pi$-pulse area, where a large spin excitation indicates low motional excitation. For durations below 60~$\mu$s, oscillatory excitation is dominant. In this regime, the excitation of the ions is extremely sensitive to the duration: Below $50 \mu$s, even qubit readout is compromised, which indicates residual energies of $\bar{n} \gtrsim 400$ phonons per ion. In this regime, exponential behavior of the energy transfer, $\bar{n} \propto \exp\left(-T/\tau\right)$, with a time constant of $\tau=$~1.4(2)$\mu$s is observed. This agrees well with the numerical simulations performed in Ref. \cite{kaufmann2014robust}. For separation durations longer than 60~$\mu$s, the excitation is mainly caused by anomalous heating. Given that the minimum total excitation is obtained if oscillatory and thermal excitation are of similar magnitude, the observed optimum value of  $\bar{n} = 4.16(0.16)$ phonons per ion is roughly consistent with the expected amount of anomalous heating of 2.3(6) phonons per ion. Rescaling the result of about two phonons per ion in 55~$\mu$s from Ref. \cite{BOWLER2012} to take into account the much lighter mass of the $^9$Be$^+$ ions used by rescaling the energy increase with $\sqrt{m}$, we achieve a rather similar energy transfer with a similar separation time.

\section{Conclusion}

We experimentally demonstrated an ion separation protocol which retains the ions in the Lamb-Dicke regime. For this goal, it is mandatory to calibrate the trap parameters accurately and to adapt the voltage ramps with respect to the findings. Best performance was achieved with separation durations comparable to that of entangling gate operations, and being well below the decoherence timescales of typical trapped ion qubit systems. We emphasize the fact that our trap geometry parameters were not specifically optimized for separation performance. Our result is consistent with the best attainable separation result, which is given by technical limitations. The lowest attained energy increase, and the energy increase at longer separation times is determined by anomalous heating. This effect is currently subject to intense research efforts, and substantial progress has already taken place. It has been demonstrated that cryogenic ion traps \cite{CHIAVERINI2014} and in-situ cleaned trap surfaces \cite{daniilidis2013probing,HITE2012} can suppress electric-field noise by several orders of magnitude, which allows for mitigating the anomalous heating effect also at very low trap frequencies and reduced trap dimensions. The latter could then lead to increased anharmonic confinement at the CP, which would in turn enable faster separation operations at low energy increase, as the tighter confinement serves to suppress undesired acceleration at the CP.
Note that our approach does not yet rely on a dedicated control-strategy \cite{chen2011optimal,palmero2013fast}, such that we achieve the limit pertaining to \textit{adiabatic} separation. Future investigations will include the applicability of techniques such as invariant-engineering approach or optimal control theory to the separation process. Technological improvements on arbitrary waveform generators \cite{BAIG2013,BOWLER2013} could also increase quartic confinement at the CP by employing larger trap voltage ranges, and improved electrical characteristics would reduce the need for filtering below $\omega_{CP}$. We expect that these upcoming technical improvements, together with the technique present in this work, will enable separation durations in the 10$\mu$s range, at energy transfers below the single phonon level.\\
Already for the current state-of-the art, it is within reach to demonstrate fundamental quantum information protocols on the few-qubit level with scalable techniques, such as distribution of entanglement and quantum teleportation over mm-distances, and implementations of the Shor factorization algorithm.

\begin{acknowledgments}
This research was funded by the Office of the Director of National Intelligence (ODNI), Intelligence Advanced Research Projects Activity (IARPA), through the Army Research Office grant W911NF-10-1-0284. All statements of fact, opinion or conclusions contained herein are those of the authors and should not be construed as representing the official views or policies of IARPA, the ODNI, or the US Government. This work was also supported by the Bundesministerium f{\"u}r Bildung und Forschung (BMBF) via IKT 2020 (QK QuOReP). We acknowledge
funding by the European Community's Seventh Framework Programme
(FP7/2007-2013) under Grant Agreement No. 323714 (EQuaM). CTS acknowledges
support from the BMBF via the Alexander von Humboldt Foundation.
\end{acknowledgments}

\section{Appendix}
\subsection{Highly excited displaced thermal states}
\label{sec:thermaldisplaced}

In Sec. \ref{sec:measscheme} we have outlined a scheme to estimate mean phonon numbers pertaining to combined oscillatory and thermal excitation from Rabi oscillation data. This requires calculating the corresponding phonon distributions, for mean phonon numbers ranging up to 10$^3$. 
The expression for this distribution Eq. \ref{eq:pn} can be evaluated by considering the phonon distribution for displaced number states \cite{ZIESEL2013}:
\bea
\vert\braket{k}{\zeta,n}\vert^2&=&e^{-\vert\zeta\vert^2} \vert\zeta\vert^{2(k+n)} n!k! \nonumber \\
&\cdot & \left\lvert\sum_{l=0}^n (-1)^l \frac{\vert\zeta\vert^{-2l}}{l!(n-l)!(k-l)!} \right\rvert^2
\label{eq:dispfockprob}
\eea

The distribution for a thermal displaced state with mean thermal phonon number $\bar{n}_{th}$ is obtained by performing a thermal average over these probabilities:
\be
p_k(\bar{n}_{th},\zeta)=\sum_{n=0}^{\infty} \frac{\bar{n}_{th}}{(\bar{n}_{th}+1)^{n+1}} \vert\braket{k}{\zeta,n}\vert^2
\ee
In practice, the summation is truncated appropriately. However, already the evaluation of Eq. \ref{eq:dispfockprob} leads to numerical problems, as for large quantum numbers, a Stirling approximation for the factorials has to be invoked. The corresponding errors, albeit of small \textit{relative} magnitude, leads to large \textit{absolute} errors of the summation in Eq. \ref{eq:dispfockprob}. A possible way to circumvent this is using an approximation introduced in Ref. \cite{saito1996}, which however holds only if both $\vert\zeta\vert^2$ and $\bar{n}_{th}$ are sufficiently large.\\
We employ a numerically convenient scheme which is precise over whole parameter range of interest, which relies on numerical thermalization of the phonon distribution of a displaced state. A thermalization process of a quantized harmonic oscillator can be modeled by a set of rate equations \cite{lamoreaux1997thermalization}:
\be
\dot{p}_n=\lambda_h\left(n p_{n-1}+(n+1) p_{n+1}-(2n+1) p_n\right),
\ee 
where the heating rate is $\dot{\bar{n}}=\lambda_h$. Defining the heating kernel 
\be
K_{n,m}=n\delta_{n,m-1}+(n+1) \delta_{n,m+1}-(2n+1) \delta_{n,m}, 
\ee
the phonon distribution for a thermal displaced state is obtained from
\be
p_n(\alpha,\bar{n}_{th})=\exp\left(\lambda_h K t\right)p_n^{(coh)}(\zeta)
\ee 
where the displaced state phonon distribution $p_n^{(coh)}=e^{-\vert\zeta\vert^2}\zeta^{2n}/n!$ is represented as a vector. The heating kernel $K=D\Lambda D^{-1}$ is a tridiagonal matrix, for which the eigenvectors $D$ and eigenvalues $\Lambda$ can be computed and stored once for a given truncation phonon number. For each sampled mean thermal phonon number $\bar{n}_{th}=\lambda_h t$, the distribution 
\be
p_n(\zeta,\bar{n}_{th})=D\exp\left(\bar{n}_{th}\Lambda\right)D^{-1}p_n^{(coh)}(\zeta)
\ee
is computed by performing three matrix-vector products. 

\subsection{Trap parameter drifts}
\label{sec:drifts}

\begin{figure}[htp]
	\centering
	\includegraphics[width=0.99\columnwidth]{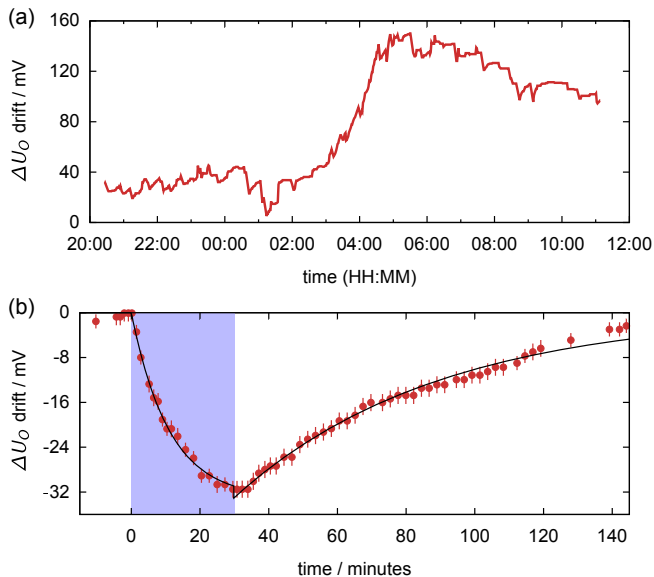}
	\caption{Drift of the tilt compensation voltage $\Delta U_O$. \textbf{a)} shows the long-time drift, caused by photoionization and superimposed thermal processes. Comparing the total magnitude of the drift with the precision required for working separation from Fig. \ref{fig:resultsplot} a), is becomes apparent that frequent, efficient and automated tilt recalibration is necessary. \textbf{b)} shows the drift caused by a single exposure of the trap to the 375~nm photoionization beam at the observation segment. The time when the laser is switched on is indicated as shaded. One can clearly recognize the different rates for charging and discharging (see text).}
	\label{fig:charging}
\end{figure}

In Sec. \ref{sec:tiltcalibration}, it was stated that the tilt force along the separation direction has to be controlled on the 100~zeptoNewton-level, corresponding to differential voltages of few mV on the $O$ electrodes, to achieve satisfactory results. This raises the question of the role of drifting trap parameters. These drifts can be caused by laser induced charging of the trap surface 
\cite{harlander2010trapped},\cite{wang2011laserinduced} acting on short (minutes) timescales, or by thermally activated surface processes on long timescales (hours). In order to use separation operations as a reliable experimental building block, the drift rates have to be kept sufficiently low in order to keep the required recalibration measurement overhead reasonable. Furthermore, an efficient measurement method for tilt calibration is required. We implement a measurement scheme similar to the technique presented in Ref. \cite{eble2010feedback}, where the ions are slowly separated under Doppler cooling illumination, the ion positions are recorded from evaluation of EMCCD images, and a possible deviation of the center-of-mass position of the ions is corrected by a digital PI servo loop with time steps of 0.5~s. This enables the determination of the tilt compensation voltage with an accuracy of 0.6~mV within a measurement time of about 5~s.\\
We find that the most substantial impact on tilt field drifts is caused by the photoionization (PI) laser beams at 423~nm and 375~nm. These either charge the trap directly via the photoelectric effect, or indirectly if ions created near the trap are accelerated onto trap surfaces by the rf electric field \cite{haerter2014longterm}. In the initial version of our setup, these lasers were sent through a common single mode fiber, and were jointly focused at the trap volume by a 150~mm achromatic lens, with an input beam diameter of about 1~mm FWHM, giving rise to foci of roughly 25~$\mu$m FWHM. However, these were found to be separated by about 50~mm along the beam direction, leading to inefficient usage of the total optical power and increased illumination of the trap surface. The beams were focused to the $C$ segments, where also Doppler cooling and imaging takes place. 
We find that with the necessity to reload ions at intervals ranging roughly between 10 minutes and 1 hour, the drifts from PI are too large to maintain good separation results. The charging rates are observed to be fluctuating and to be extremely sensitive on the beam alignment. We mitigate this problem by the following measures: 
First, we direct the 375~nm and 423~nm  beams via separate optical fibers and via separate focusing lenses. This way, we can guarantee that both foci are located at the same position along the propagation direction, which further improves the beam clearance and yields more efficient usage of the optical power. Second, the PI beams are adjusted to a position 5 segments away from the $C$ segment, such that charging takes place at locations with reduced electric feed-through at $x=0$. We realize a remote loading procedure by periodically shuttling a confining axial potential well from the PI to the observation site within 160~$\mu$s.\\
These measures enable us to achieve sufficiently low drift rates. We quantify these by recording the tilt compensation voltage $\Delta U_O$ as determined by calibration measurements over time. Fig. \ref{fig:charging} shows the results for the case of a single PI event and a long-time record. For the first case, the charging/discharging process is modeled as follows \cite{wang2011laserinduced}:
\be
\dot{Q}=K-\delta Q-\kappa Q
\label{eq:charging}
\ee
where $Q$ is the charge accumulated on insulating patches, to which the tilt field $\gamma'$ from Eq. \ref{eq:gammadef} is assumed to be proportional, $\gamma'=c\cdot Q$. $K$ is the laser induced charging rate, $\delta$ describes the screening due to the existing charge, and $\kappa$ is the discharge rate. $K$ and $\delta$ are only nonzero when the PI lasers are on. As tilt compensation is given with $\Delta U_O \gamma_O=\gamma'$, we can measure the quantity $K'=c\cdot K/\gamma_O$ by observing the change of the tilt compensation voltage under exposure of a PI beam by fitting the data to the model Eq. \ref{eq:charging}. For the 375~nm PI laser at 400$\mu$W and the remote setting, we determine $K'=$3.02(6)mV/min, $\delta=$0.074(2)~min$^{-1}$ and $\kappa=$0.017(2)~min$^{-1}$. No significant charging induced by the PI laser near 423~nm at 400~$\mu$W could be observed for the remote setting. By contrast, for direct loading at the $C$ segment, we measure rates of change for the tilt compensation voltage of $K'=$4.8(2)mV/min (375~nm) and $K'=$5.25(2)mV/min (423~nm). The measurement data presented in the main text was acquired with remote loading and PI powers of about 300~$\mu$W (1~mW) in the 375~nm (423~nm) beam. The rather large powers are chosen because we presume the trap loading rate to be proportional to the laser power, and for the typical loading time of about 5~s we obtain, we are in the regime where the accumulated charge is proportional to the time the PI lasers are switched on, as indicated by Fig. \ref{fig:charging} b). We thus assume that for lower PI powers, the longer required loading times would give rise to similar total changes of the tilt compensation voltage.

\bibliography{splittingExp}

\end{document}